\documentclass[fleqn,usenatbib]{mnras}

\usepackage{newtxtext,newtxmath}
\usepackage{verbatim}

\usepackage[T1]{fontenc}

\DeclareRobustCommand{\VAN}[3]{#2}
\let\VANthebibliography\thebibliography
\def\thebibliography{\DeclareRobustCommand{\VAN}[3]{##3}\VANthebibliography}

\usepackage{graphicx}	
\usepackage{amsmath}	

\newcommand{\fermi}{{\it Fermi}-LAT}
\newcommand{\src}{S255 NIRS 3}
\newcommand{\latsrc}{4FGL\,J0613.1+1749c}

\title[High energy emission from S255 NIRS 3]{High energy gamma-ray emission powered by a young protostar: the case of S255 NIRS 3}%

\author[]{Emma de O\~na Wilhelmi$^{1}$\thanks{E-mail: Emma.de.ona.wilhelmi@desy.de}, Rub\'en L\'opez-Coto$^{2}$, Yang Su$^{3}$
\\
$^{1}$ Deutsches Elektronen-Synchrotron DESY, Platanenallee 6, 15738 Zeuthen, Germany\\
$^{2}$ Instituto de Astrof\'isica de Andaluc\'ia, CSIC, 18080 Granada, Spain.\\
$^{3}$Purple Mountain Observatory and Key Laboratory of Radio Astronomy, Chinese Academy of
Sciences, Nanjing 210034, China\\
}

\date{Accepted XXX. Received YYY; in original form ZZZ}

\pubyear{}

\begin{document}
\label{firstpage}
\pagerange{\pageref{firstpage}--\pageref{lastpage}}
\maketitle

\begin{abstract}
Evidence of efficient acceleration of cosmic rays in massive young stellar objects has been recently reported. Among these massive protostars, \src\ for which extreme flaring events associated with radio jets have been detected, is one of the best objects to test this hypothesis. We search for gamma-ray emission associated with this object in {\it Fermi}-LAT data and inspect the gas content in different molecular lines using the MWISP survey. A GeV source dubbed \latsrc\ lies on top of the MYSO region, where two filamentary $\sim$10\,pc CO structures extend along the same direction of the sub-parsec radio jets. We investigate the spectrum, morphology, and light curve of the gamma-ray source and compare it with the theoretical emission expected from hadronic and leptonic populations accelerated in the radio jets. We argue that the gamma-ray source could be powered by particles accelerated in the \src\ jets, radiating via Bremsstrahlung or proton-proton interaction, and with a synchrotron component shinning in radio from primary or secondary electrons in the case of a leptonic or hadronic population. 
\end{abstract}




\section{Introduction}

The recent discoveries of high energy gamma-ray radiation from stellar winds, such as novae or stellar clusters have demonstrated the efficient acceleration of particles in shocks associated with stellar objects \citep{2022ApJ...935...44C,2022Sci...376...77A,2014Sci...345..554A,2022NatAs...6..689A,2019NatAs...3..561A,2010Sci...329..817A}. Related to them, massive young stellar objects (MYSOs) have also been hypothesized as cosmic-ray accelerators, as they power fast outflows associated to accretion mechanisms or supersonic ejecta \citep{2021MNRAS.504.2405A,2007A&A...476.1289A,2015A&A...582L..13P,2010A&A...511A...8B}. The observation of non-thermal emission from jets associated with some of these objects indicates the presence of relativistic electrons \citep{2003ApJ...594L.131B,2003ApJ...587..739G,2010Sci...330.1209C,1993ApJ...416..208M}. These outflows appear at linear distances of less than a parsec from the MYSO, where the environment is strongly affected by its dense IR photon field and high-density matter from the hosting cloud. The population of relativistic particles might potentially radiate in gamma rays via leptonic (inverse Compton and Bremsstrahlung) and/or hadronic (proton-proton interaction) processes. 

S255 NIRS 3 (or G192.60--00.04) is a well-studied $\sim$20 M$_\odot$  high-mass young stellar object (HMYSO) in the S255IR massive star-forming region, located at a distance
of $\sim$1.8~kpc \citep{2016MNRAS.460..283B}. It exhibits a disk-like rotating structure associated with an accretion disk. A molecular bipolar outflow or jet-like outflow (with a size of $\sim1$~arcmin) has also been detected perpendicular to the disk at position angle 67$^{\rm{o}}$ \citep{2015ApJ...810...10Z}. Several measurements indicate variability of such jet, pointing to irregular outburst episodes, that occur every few hundred years \citep{2018A&A...612A.103C,2016MNRAS.460..283B}. This type of massive star is believed to be formed through disk-mediated accretion. \src\ gained appreciable attention for its extreme flaring signatures in 6.7 GHz methanol
maser, IR, and radio continuum \citep{2018A&A...617A..80S,2016MNRAS.460..283B,2016ATel.8732....1S,2017NatPh..13..276C,2018A&A...612A.103C}. The detection of a sudden increase in the FIR stellar bolometric luminosity, from 2.9$\times10^4$L$_{\odot}$ to 1.6$\times10^5$L$_{\odot}$ at around 30-100 $\mu$m band, was invoked to demonstrate the irregularity of the accretion flow, rather than a smooth, steady one. These accretion events are believed to trigger an increase in the mass ejection rate, which results in an enhancement of the photon field in the surroundings of the star and eventually a jet ejection. 

Motivated by this discovery, we explore the high energy emission using data from the LAT telescope on board the {\it Fermi} satellite. A source dubbed 4FGL\,J0613.1+1749c\footnote{The {\it c} assigned to the name designs a region coincident with interstellar clump} listed in the \fermi\ Fourth Source Catalog (4FGL-DR3, \citealt{2020ApJS..247...33A}) lies at RA$_{\rm J2000}$=93$^{\rm o}$.28, DEC$_{\rm J2000}$=17$^{\rm o}$.822, coincident with the position of the MYSO source.  

In the following, we investigate the connection between the LAT source with \src, and discuss the implications of such a connection, under the hypothesis of a non-thermal jet powered by the protostar.

\section{Data Analysis and Results}

\begin{figure}
  \centering
  \includegraphics[width=0.48\textwidth]{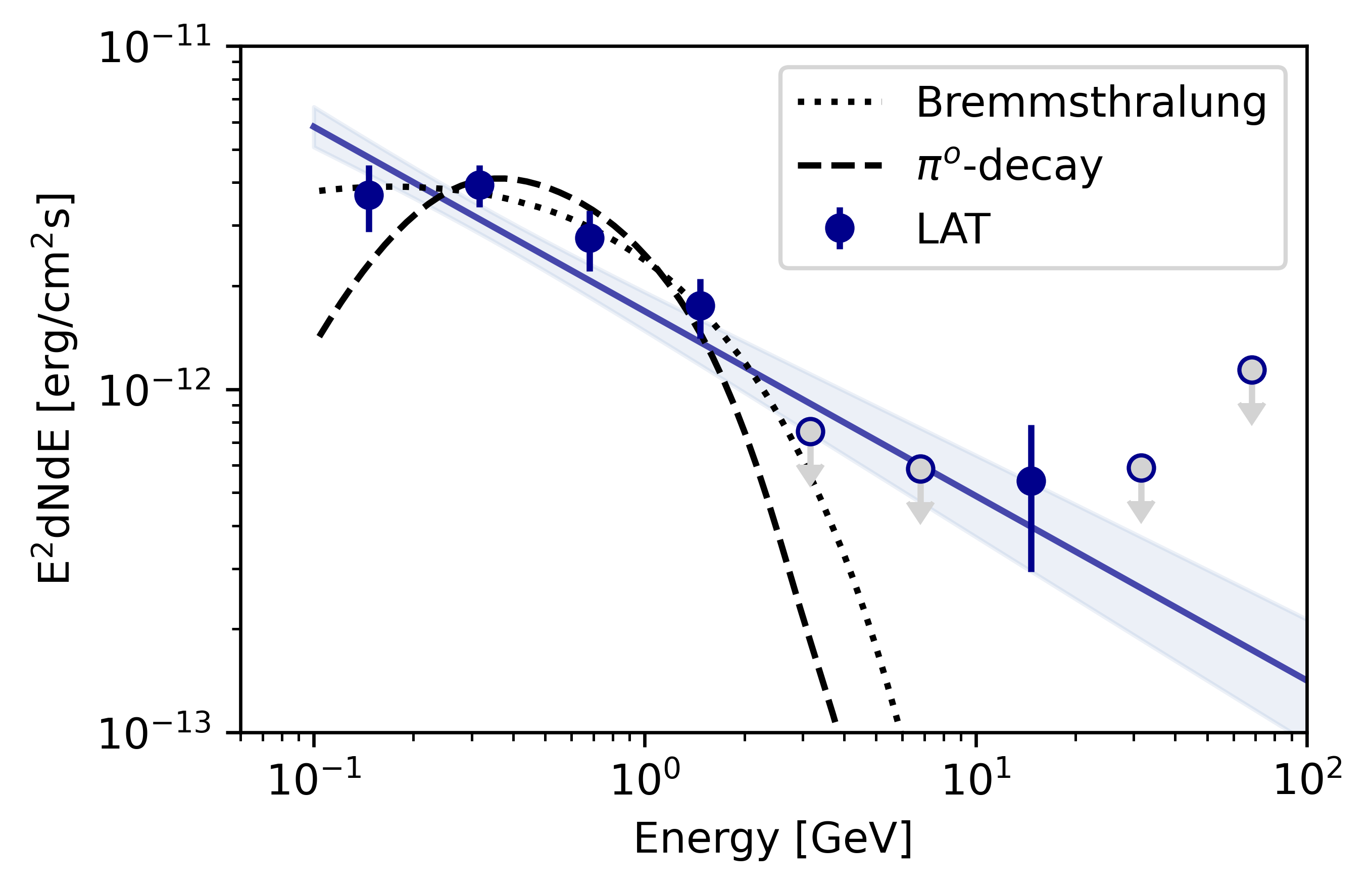}

  \caption{Spectral energy distribution derived from 4FGL J0613.1+1749c. The 1-$\sigma$ error in the best-fit parameters is shown with a shaded blue area. The dashed and dotted lines show the best-fit models obtained for hadronic and leptonic emission mechanisms. }
  \label{fig:spec}
\end{figure}

\subsection{Fermi-LAT analysis}

To investigate the nature of the gamma-ray source located at the position of the MYSO source, we obtained \fermi\ data from 4$^{\rm th}$ August 2008 to 25$^{\rm th}$ January 2023. The \fermi\ (P8R3, pass 8 release 3) data were analyzed in the energy range from 100\,MeV to 600\,GeV, around the region of interest (ROI), which was defined by a radius of $20^\circ$ around the position of \src. The analysis of the LAT data described below is performed by means of the $\textsc{fermipy}$ $\textsc{python}$ package (version 0.18.0), based on the $\textsc{Fermi Science Tools}$ (version 2.0.8, \citealt{2017ICRC...35..824W,2019ascl.soft05011F}). The SOURCE class events with a maximum zenith angle of
$90^\circ$ were selected to eliminate Earth limb events. To derive the global spectrum, the data are binned in 8 energy bins per decade and spatial bins of $0.1^\circ$ size. The response of the LAT instrument is evaluated with the IRFs (version P8R3\_SOURCE\_V3). The gamma-ray emission in the ROI was described using all LAT sources listed in the \fermi\ 4FGL-DR3 catalog in a radius of $25^\circ$ around the position of the MYSO. The contributions from Galactic and extra-galactic diffuse gamma-rays are modeled using the Galactic (gll\_iem\_v07) and isotropic (iso\_P8R3\_SOURCE\_V3\_v1) diffuse emission models, available from the Fermi Science Support Center (FSSC) \footnote{http://fermi.gsfc.nasa.gov/ssc/data/access/lat/BackgroundModels.html}. The energy dispersion correction is applied to all sources in our model, except for the isotropic diffuse emission model. Additional gamma-ray emission regions which might not be accounted for in the 4FGL catalog were investigated using the \textsc{findsource} method, applying a criterium of Test Statistics TS$>$25, resulting in one additional source which was added to the model (located at (RA, DEC)$_{\rm J2000}$=(90.88,18.34)$^{\rm o}$). The inclusion of this source in the model had no effect on the fit results of the source of interest. The spectral parameters of the sources within $4^\circ$ of \src, sources with TS larger than 500, and the Galactic and isotropic
diffuse emission components were left free in the likelihood fit. To compute the lightcurve, the data set was split into 10 bins using the \textsc{lightcurve} method in \textsc{fermipy}. 

\begin{figure}
  \centering
   \includegraphics[width=0.45\textwidth]{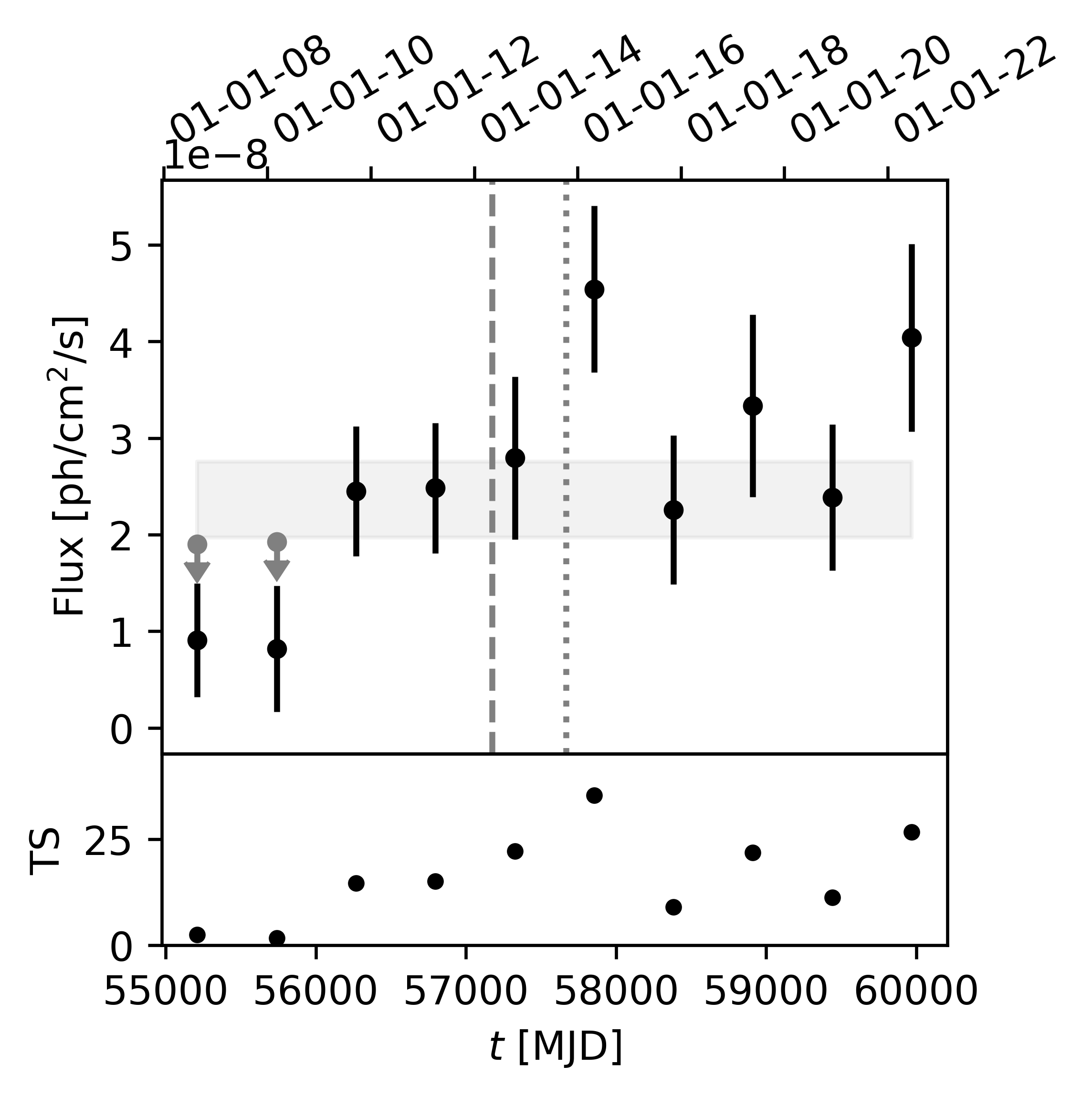}
  \caption{Gamma-ray light-curve of the observations ranging 14 years. The dashed vertical line shows the beginning of the maser flare reported by \protect\cite{2015ATel.8286....1F} whereas the dotted line marks the beginning of the radio flare reported by \protect\cite{2018A&A...612A.103C}. The shaded area shows the 1$\sigma$ error on the photon flux integrated over the analysis energy range.
  }

  \label{fig:lc}
\end{figure}

The spectral and morphological results from the fit for the GeV source coincident with \src\ are compatible with the ones in the 4FGL catalog: \latsrc\ is well-described (with test-statistics of TS=113.6) by a point-like source with a flux well-described by a power-law spectral function of the form $\phi(E) = \phi_0 \left( \frac{E}{E_{o}}\right)^{-\alpha}$, with $\phi_0=(6.21 \pm  0.73)\times 10^{-13}$ MeV$^{-1}$cm$^{-2}$s$^{-1}$ and $\alpha=2.54 \pm 0.07$ (E$_{\rm o}$ fixed to 1.231~GeV). The spectral energy distribution is shown in Fig. \ref{fig:spec}. To test the extension of the source, we selected only events with energy above 3\,GeV to optimize the LAT angular resolution\footnote{https://www.slac.stanford.edu/exp/glast/groups/canda/lat\_Performance.htm}. We obtain an upper limit (95 percent CL) on the extension of 0.25$^{\rm{o}}$.

The localization of the point-like source results in a best-fit position at RA$_{\rm J2000}$ =(93.27$\pm$0.03)$^{\rm o}$ and DEC$_{\rm 2000}$ = (17.96$\pm$0.04)$^{\rm o}$, well in agreement with the position of the MYSO source. The contours above 12$\sigma$ obtained from the skymap produced for events with energy higher than 1\,GeV (selected as a compromise between enough statistics and angular resolution) is shown in Fig. \ref{fig:skymap1GeV} in white. 

The light curve was derived by dividing the data set into 10 equal bins in time. We chose 10 bins as a good compromise between time resolution and detection significance. Figure \ref{fig:lc} shows the integral flux for each time bin on the upper panel, and the TS on the lower one. An upper limit on the flux (95 percent CL) is displayed if the test statistics (TS) is less than 4. The two vertical lines mark the start of the 6.7 GHz methanol maser reported by \cite{2015ATel.8286....1F} and the beginning of the radio flare reported by \cite{2018A&A...612A.103C} (dashed and dotted lines, respectively). The gamma-ray source is detected well above 20 TS at the moment of the radio outburst, whereas before 2016 the flux level seems to be very low. An indication for flux variability is found at a 2.7${\sigma}$ level ($\chi^2$/ndf=22.77/9). However, at this significance, this variability might well be due to statistical fluctuations and no firm statements can be made.

\subsection{CO observations}

\label{sec:co}  

The molecular complex in this region has been studied in detail in several lines and scales \citep{1985MNRAS.216..713R,2008ApJ...682..445C,2009AJ....138..975B,2012ApJ...755..177Z,2015ApJ...810...10Z,2016MNRAS.460..283B,2022PASJ...74..545K}. We investigated the large-scale CO emission on the surrounding medium of \src\ using deep observations in the region from the MWISP survey (see details in \citealt{2019ApJS..240....9S}). The $^{12}$CO(J=1--0), $^{13}$CO(J=1--0), and C$^{18}$O (J=1--0) lines were simultaneously observed using the 13.7\,m telescope. The covered region is towards the northern Galactic plane with $|b|< 5^{\rm o}$. The half-power beam width (HPBW) is $\sim$50 arcsec at the frequency of $\sim110-115\,\rm{GHz}$. The rms noise is $\sim0.5\,$K for $^{12}$CO and $\sim$0.3\,K for $^{13}$CO and C$^{18}$O, at a velocity resolution of $\sim$0.2\,km s$^{-1}$ with a uniform grid spacing of 30 arcsec. More details about these observations and a very detailed analysis of the dense cloud around \src\ can be found in \cite{2017ApJS..230....5W}. The data cubes were analyzed using the \textsc{CARTA} image visualization and analysis tool \citep{2020zndo...3746095C}. 

\begin{figure}
  \centering
   \includegraphics[width=0.45\textwidth]{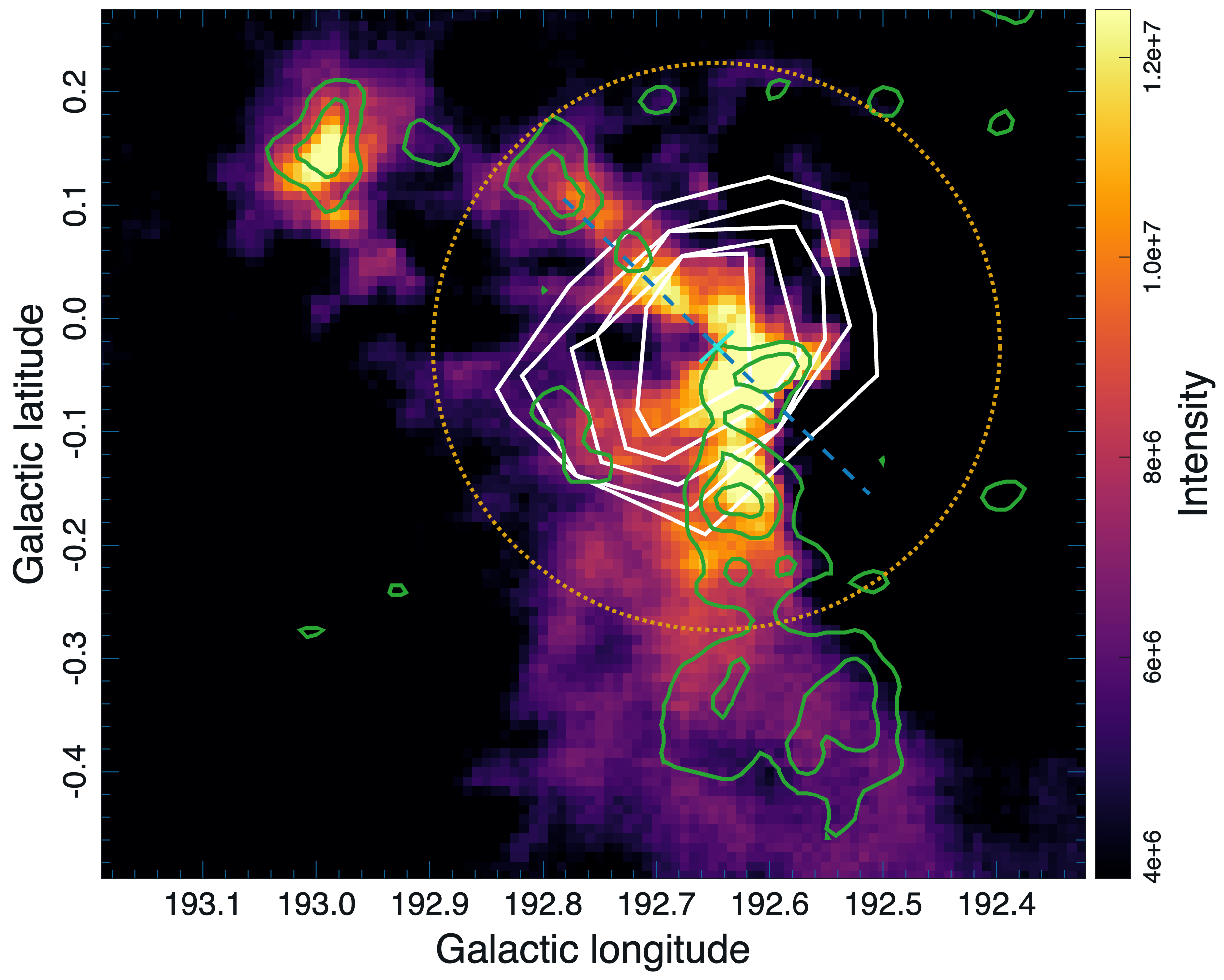}
  \caption{Integrated $^{13}$CO intensity map of the \src\ region obtained from the MWISP survey. The green contours are obtained from the C$^{18}$O intensity map, tracing the two extense filamentary structures emerging from the core of the emission. The contours above 12 TS (in steps of 2 TS) obtained from the LAT Test-Statistic skymap above an energy threshold of 1\,GeV are overlaid in white and the best-fit position in cyan. The direction of the 1 arcmin molecular bipolar outflow is shown with a blue dashed line. The orange circle marks the 95 percent CL on the source extension.
  }
  \label{fig:skymap1GeV}
\end{figure}

The velocity profile of the region around \src\ is dominated for a prominent single peak at v$_{\rm LST}\sim$7.5\,km~s$^{-1}$, which is close to the H$_2$O maser's velocity of ~6.4\,km~s$^{-1}$ for \src\ (see Figures 1-2 in \citealt{2021A&A...647A..23H}). We thus assume that the dense molecular cloud is at a distance of 1.78 kpc \citep{2016MNRAS.460..283B} based on the physical association between the filamentary cloud and the masers. The CO intensity is dominated by the stellar complex S255. A high-density peak can be found at the position of the cluster at a velocity range between 4\,km~s$^{-1}$ and 11\,km~s$^{-1}$. In Fig. \ref{fig:skymap1GeV} we show the $^{13}$CO image integrated in this velocity range. Overlaid with green contours we plotted the integrated intensity of C$^{18}$O (see also Fig. 8 in \citealt{2017ApJS..230....5W}). The distribution of significance of the GeV emission is peaked in the central region of \src, where the maximum of the CO emission is observed. Interestingly, the two distinct filamentary structures reported by \citealt{2017ApJS..230....5W, 2009AJ....138..975B} are aligned with an angle of $\sim$65$^{\rm{o}}$ with the jet structure reported by several authors \citep{2011A&A...527A..32W,2015ApJ...810...10Z}, but at much larger scale ($\sim$10\,pc) (see Fig. \ref{fig:skymap1GeV}, dashed blue line).

The molecular gas mass directly related to the \src\ (the dust lane or the dense gas filament at V$_{\rm LSR}$=(4-11)\,km~s$^{-1}$) is about $10^3$M$_{\odot}$ (in a region of 17 arcmin length and 4 arcmin width, or 8.8$\times$ 2.1 pc$^2$), while we estimate a total molecular gas mass of about 10$^4$M$_{\odot}$ in the region where the gamma-ray source sits (see white contours on Fig. \ref{fig:skymap1GeV}). The mean density of the molecular gas in the filament is about 300-400\,H$_2$ cm$^{-3}$ or even higher. These rich and dense region provides sufficient target to promote gamma-ray emission via Bremsstrahlung or proton-proton interaction if relativistic electrons and/or protons are accelerated somewhere in the region.

\section{Discussion}

The rich multi-wavelength data around the MYSO source \src\ allows us to investigate the possibility of gamma-ray high energy emission associated with accretion-ejection phenomena in the massive protostar. 
This particular MYSO, with a (steady) luminosity of 5$\times10^{4}$L$_{\odot}$, is one of the few objects that has experienced rare but intense outbursts of radio emission, in which an increase in the luminosity up to 1.3$\times10^{5}$L$_{\odot}$ was measured \citep{2017NatPh..13..276C}. These outbursts are believed to be recurrent and are associated with phenomena related to disk accretion and the ejection of a jet. In 2015 a flare of the Class II methanol maser was reported \citep{2015ATel.8286....1F}. Follow-up mid-infrared and far-infrared observations in November 2015 showed a sudden increase in the brightness of the central object. Later in July 2016 (13 months later from the maser outburst), a radio jet burst triggered by the accretion burst was also detected. This outburst released an estimated energy of 1.2$\times10^{46}$erg (calculated from the beginning of the burst until mid-April 2016, \citealt{2017NatPh..13..276C}). Using high-resolution Jansky Very Large Array (JVLA) observations and comparing observations from 2012 and 2018, \cite{2021MNRAS.501.5197O} reported the presence of two non-thermal radio lobes associated with the protostar, with energy index (S$_\nu\propto\nu^{-\alpha}$) of $\alpha\simeq0.6$, corresponding to an electron index of p = 2.2. The latest is in good agreement with what is expected from an uncooled population of relativistic electrons produced by diffusive shock acceleration (DSA) at a strong non-relativistic shock (e.g. \citealt{1999tcra.conf..247P}). 
These lobes are located at a projected distance of 0.02 and 0.04~pc (or $6.2\times10^{16}$ and $1.2\times10^{17}$~cm, respectively) from the compact center. They are also aligned in a direction similar (with an angle of $\approx$67$^{\rm o}$) to the orientation of the radio emission of the compact center, the infrared emission in the field, and the observed CO filaments (see Fig. \ref{fig:skymap1GeV}) pointing to a jet-like emission powered by the central object. Based on equipartition considerations, \cite{2021MNRAS.501.5197O} estimated a minimum energy and a minimum magnetic field strength of $4\times10^{41}$~erg and 1.1\,mG, respectively. These radio lobes are probably generated by the strong terminal shocks off the jets, which have an estimated velocity of $\sim$700 km~s$^{-1}$ \citep{2021MNRAS.501.5197O,2018A&A...612A.103C}. This value is within the range of jet velocity expected from jets in these objects (see \citealt{1993ApJ...416..208M,1995ApJ...449..184M}).

Unfortunately, the lightcurve cannot establish an unambiguous link between the observed flare and the gamma-ray emission reported here. Even if the increase of the significance and flux towards the time of the flare makes this possibility attractive, the statistical analysis of the data shows no significant variability further than 2.7$\sigma$ (without accounting for trials and systematic errors), preventing such a connection. 

However, the presence of a shock in which relativistic electrons are accelerated and the amplified magnetic field, in a dense region, suggests that the origin of \latsrc\ should be related somehow to \src. The potential of massive young objects to power gamma-ray sources was investigated by several authors (see e.g. \citealt{2010A&A...511A...8B,2007A&A...476.1289A,2021MNRAS.504.2405A}), and applied to several MYSO with similar parameters to the ones found in \src. We follow thus the approach described in \cite{2010A&A...511A...8B} to explore the expected gamma-ray emission in case of electrons and protons accelerated in jets associated with \src, and compare with the observations described above.

\begin{figure}
  \centering
  \includegraphics[width=0.45\textwidth]{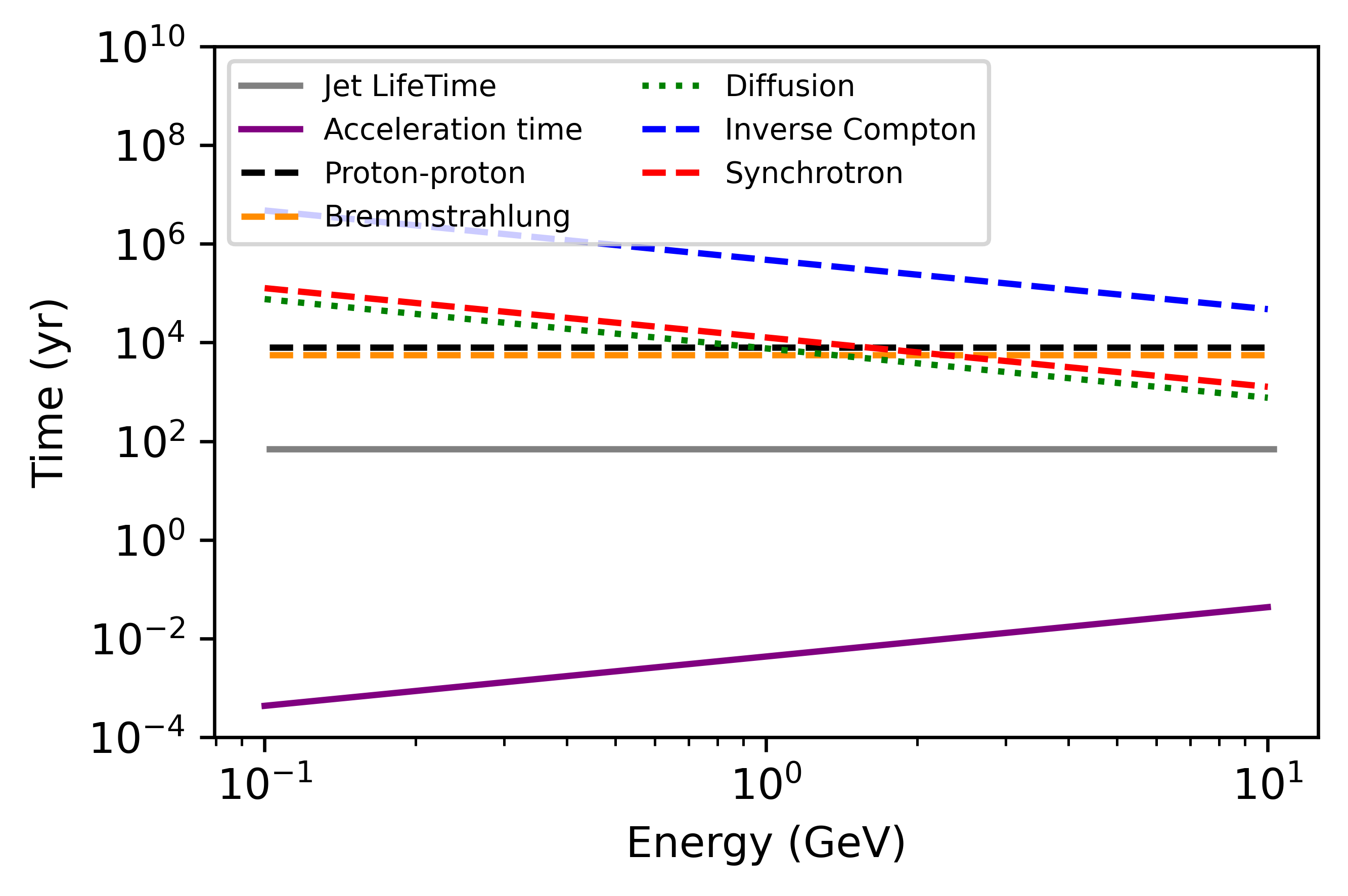}

  \caption{Relevant timescales as a function of the particle energy applied to the massive young stellar object \src.}
  \label{fig:coolingtimes}
\end{figure}

\src\ is located at a distance of 1.8 kpc and the total luminosity above 100\,MeV amounts L$_{\gamma} = (4.2 \pm 0.5)\times10^{33}$ erg~s$^{-1}$. Relativistic particles, hadrons or leptons, can be accelerated in jets, even for moderated flow speed \citep{2022Sci...376...77H,2022NatAs...6..689A}. For electrons, the relevant mechanisms ruling the observed spectral energy distribution are synchrotron, inverse Compton, and relativistic Bremsstrahlung, while for protons, inelastic collision via proton-proton would originate the observed gamma-ray emission, whereas the radio emission would correspond to synchrotron emission from secondary electrons. We used the estimated magnetic field of 1\,mG, which is also in agreement with the typical magnetic field obtained from Zeeman measurement in dense clouds like the one in \src\ (see Section \ref{sec:co}). Relativistic electrons can upscatter the dense FIR photon field in the region to the GeV regime through inverse Compton. We calculated the IR photon density in the region of the jet using the one reported during the flare ($1.6\times10^5 L_{\sun}$), to account for the maximum-possible contribution. For radiation mechanisms depending on the gas density (i.e. Bremsstrahlung and proton-proton collisions), we used a minimum gas density at the location of the jets of n$_{\rm H}\simeq200$~cm$^{-3}$, which is also in agreement with the values of the jet density obtained in \cite{2010A&A...511A...8B}.  

In Fig. \ref{fig:coolingtimes} we show the cooling times for the mechanisms described above, as well as the relevant times for the shock evolution in which particles are believed to be accelerated such as the jet lifetime, diffusive time for particles in the Bohm limit and acceleration time, for a jet of velocity and magnetic field of 700~km~s$^{-1}$ and 1\,mG. The dominant gamma-ray emission processes are proton-proton interaction for hadrons and Bremsstrahlung in electrons (black and orange dashed lines, respectively) with the maximum energy determined by the acceleration time in the jets (purple solid line). For inverse Compton to dominate over Bremsstrahlung, the density has to be lower, requiring a larger emission region of the order of a few parsecs. However, in such a larger region, the IR photon density is also greatly suppressed, resulting in a negligible contribution of the inverse Compton component. This larger region where the large-scale CO filamentary structure is found is still confined within the angular extension for a point source in LAT. Alternatively to the emission region located on the jet shock region, recurrent shocks could have injected particles that diffuse out into a region of $\sim$10~pc around the protostar. However given the large density in the sub-parsec region around \src\ only high-energy particles would escape without losing their energies via Bremsstrahlung/proton-proton radiation, rendering this scenario less likely.

If the gamma-ray emission is indeed associated with the non-thermal jets observed in \src, we can obtain the electron and/or proton population behind the radio and GeV emission, by fitting the spectral energy distribution using the \textsc{Naima} library within the framework of the \textsc{Gammapy} software \citep{2015ICRC...34..922Z,gammapy:2017}. In both cases, the particle spectrum was described with a power-law function with an exponential cutoff in energy. A single population of electrons radiating via synchrotron and Bremsstrahlung in the radio and GeV energy range, respectively, can in principle reproduce the spectral shape measured. The spectral index of the radio emission in the jets constrains the local electron power-law index, which is compatible with the best-fit particle spectral index obtained ($\rm{p}=1.8\pm0.7$), with an unconstrained energy cutoff at $\rm{E_{c,e}}\simeq1.9$~GeV (see dotted line in Fig. \ref{fig:spec}). Using the expression in \citealt{2010A&A...511A...8B}, for primary electrons:

\begin{equation}
B\sim 0.04(L_{35}^{\rm e})^{-2/3}R_{\rm 16}^{-2/3}\nu_{5 GHz}^{1/3}d_{\rm 3~kpc}^{4/3}~F_{\rm \nu~mJy}^{2/3}~{\rm mG},
\end{equation}
with $L_{35}^{\rm e}$ the luminosity of the electron population in erg~s$^{-1}$, R the size of the jets in units of $10^{16}$cm, and $F_{\rm \nu}$ the measured radio flux at $\nu$=5~GHz ($\sim$0.65\,mJy, see \citealt{2021MNRAS.501.5197O}), results on a magnetic field strength of less than 0.1~mG, which is below the minimum magnetic field estimated in the region. 

The fit to a proton-proton distribution with particle spectral index fixed to 2 for the uncooled energy spectrum, as expected by DSA in shocks (e.g. \citealt{1983RPPh...46..973D}), reproduces the GeV data correctly with an energy cutoff of $\rm{E_{c,p}}=3.1\pm0.9$~GeV (see dashed lines in Fig. \ref{fig:spec}). The radio lobes can in the scenario be attributed to synchrotron emission from the secondary electrons produced in the GeV hadronic emission. By comparing the radio and GeV flux (using the expression above, and substituting $L^e_{35}$ by its equivalent for secondary electrons, 0.1(t$_{\rm esc}$/t$_{\rm pp})L^p_{35}$) a magnetic field of 1~mG can be estimated, which agrees well with the magnetic field obtained in \citealt{2021MNRAS.501.5197O}, and also with the magnetic field strength considered in theoretical models. The hadronic origin fits well the data in a one-zone scenario, however, it should be noted that a leptonic origin cannot be discarded, given that the lower magnetic field required by the synchrotron emission can be compensated by a moderate increase of the density on the emission region. In such a scenario though, a departure from the equipartition hypothesis has to be assumed.

\section{Conclusions}

We investigated the possibility of GeV gamma-ray emission related to the accretion-ejection phenomena in the massive protostar \src. We found a point-like gamma-ray source, dubbed in the 4FGL catalog as \latsrc, located at a position compatible with the one of the protostar. 
In general, associating MYSO objects with gamma-ray sources in terms of positional coincident is difficult, given the ubiquity of these kinds of objects in massive clouds, which do correlate with the gamma-ray distribution of sources in the Galaxy \citep{2014A&A...565A.118P,2011A&A...530A..72M}. Additionally, several massive OB stars and two non-thermal X-ray sources associated with background AGNs, are located within the region of the gamma-ray source \citep{2011A&A...533A.121M}, adding uncertainty to the identification of the gamma-ray source with the MYSO. Similar massive stars are found in high-energy binary systems such e.g. LS\,5039 \citep{1997A&A...323..853M}, however, no compact object has been found associated with these ones in the S255 complex. No other types of sources usually associated with Galactic GeV gamma-ray radiation, such as pulsars or supernova remnants are found either.

The location of \src\ close to the anticenter, where the stellar formation regions are not so common, the lack of any other obvious counterpart, and the point-like nature of the gamma-ray source 4FGL\,J0613.1+1749c make the association with \src\ appealing. Additionally, the observed spectral energy distribution in radio and GeV wavelengths is in good agreement with the emission expected theoretically, when applied to the specific conditions of this object. The molecular-rich environment provides enough targets to explain the gamma-ray emission proton-proton interactions, with protons accelerated in the observed jet (or in recurrent ones). The estimated magnetic field in such jets is also in agreement with the one required to explain the radio emission by synchrotron emission from secondary electrons. 

The CO image also unveiled a large-scale molecular filament, oriented in the same direction as the radio compact jets, which could be originated from recurrent ejections powered by the protostar. 

A correlation through the lightcurve between the MWL flare and the gamma-ray emission would leave no doubt about the origin of the gamma-ray source. However, the hint of variability is too low to claim the flare as powering the gamma-ray source. Two more of these sources with similar environments, that is, massive young protostars with jets, have been also proposed as counterparts of gamma-ray sources, namely HH 211 and HH 80--81 \citep{2023NatAs.tmp...15Y,2022RAA....22b5016Y}.

Monitoring of this kind of flares should clarify if the observed gamma-ray radiation can be univocally attributed the MYSO activities, probing the role of cosmic rays in the evolution of massive young stellar objects.

\section*{Acknowledgments}
 EdOW acknowledges the support of DESY (Zeuthen), a member of the Helmholtz Association HGF. R.L.-C. acknowledges the Ram\'on y Cajal program through grant RYC-2020-028639-I and the financial support from the grant CEX2021-001131-S funded by MCIN/AEI/ 10.13039/501100011033. We thank Rolf B\"uhler for useful comments on the LAT analysis and Felix Aharonian, Brian Reville and Anabella Araudo for helpful discussions on the interpretation. We also thank to Alessio Caratti o Garatti for his insights on the physics of this object.
 This research made use of the data from the Milky Way Imaging Scroll Painting (MWISP) project, which is a multiline survey in $^{12}$CO/$^{13}$CO/C$^{18}$O along the northern Galactic plane with the PMO 13.7m telescope. We are grateful to all the members of the MWISP working group, particularly the staff members at the PMO 13.7m telescope, for their long-term support. MWISP was sponsored by the National Key R\&D Program of China with grant 2017YFA0402700 and the CAS Key Research Program of Frontier Sciences with grant QYZDJ-SSW-SLH047.

\section*{Data Availability}
The Fermi data underlying this article are available at https://fermi.gsfc.nasa.gov/ssc/data/access/lat/. The CO data was accessed from the MWISP survey and can be shared on reasonable request to the corresponding author. This publication uses data products from the Wide-field Infrared Survey Explorer, a joint project of the University of California, Los Angeles, and the Jet Propulsion Laboratory/California Institute of Technology, funded by the National Aeronautics and Space Administration.



\bibliographystyle{mnras}
\bibliography{biblio} 




\bsp	
\label{lastpage}
\end{document}